\documentclass[%
superscriptaddress,
amsmath,amssymb,
aps,
twocolumn,
prl,
floatfix,
]{revtex4-2}
\usepackage{siunitx}
\usepackage[english]{babel}  % Add this line
\usepackage{graphicx}% Include figure files
\usepackage{dcolumn}% Align table columns on decimal point
\usepackage{bm}% bold math
\usepackage{soul}
\usepackage{xcolor}
\usepackage{float} % to force figures to stay at location 
\renewcommand{\selectlanguage}[1]{}  % ignore language - {en} in bibfile, https://tex.stackexchange.com/questions/508521/revtex-4-2-package-babel-error-you-havent-defined-the-language-en-yet
\renewcommand*{\vec}[1]{\bm{#1}}

\begin{document}

\preprint{APS/123-QED}

\title{Nature of granular drag in microgravity}

\author{Tivadar Pong\'o}
\thanks{These authors contributed equally to this work.}
\affiliation{%
 Collective Dynamics Lab, Division of Natural and Applied Sciences, Duke Kunshan University, 215306 Kunshan, Jiangsu, China}%

\author{Tianhui Liao}%
\thanks{These authors contributed equally to this work.}
\affiliation{%
 Collective Dynamics Lab, Division of Natural and Applied Sciences, Duke Kunshan University, 215306 Kunshan, Jiangsu, China}%
\affiliation{School of Science, Westlake University, Hangzhou 310030, Zhejiang, China}

\author{Jinchen Zhao}%
\affiliation{%
 Collective Dynamics Lab, Division of Natural and Applied Sciences, Duke Kunshan University, 215306 Kunshan, Jiangsu, China}%

\author{Valentin Dichtl}%
\affiliation{%
Experimentalphysik V, Universit\"at Bayreuth, 95440 Bayreuth, Germany}%

\author{Simeon V\"olkel}%
\affiliation{%
Experimentalphysik V, Universit\"at Bayreuth, 95440 Bayreuth, Germany}%

\author{Ra\'ul Cruz Hidalgo}%
\affiliation{Department of Physics and Applied Mathematics, University of Navarra, 31009 Pamplona, Spain}

\author{Kai Huang}%
 \email{kai.huang186@dukekunshan.edu.cn}
\affiliation{%
 Collective Dynamics Lab, Division of Natural and Applied Sciences, Duke Kunshan University, 215306 Kunshan, Jiangsu, China}%
 \affiliation{Experimentalphysik V, Universit\"at Bayreuth, 95440 Bayreuth, Germany}%Lines break

%\\
% \textbackslash\textbackslash
 
\begin{abstract}

The influence of gravity on the drag force acting on a projectile impacting granular media is investigated experimentally via embedded inertial measurement unit (IMU) sensor and numerically through discrete element method (DEM) simulations. As gravity approaches zero, inertial drag dominates, yielding qualitatively different scaling laws and cavity dynamics. Analogous to fluid dynamics, we define a dimensionless granular drag coefficient $C_{\rm gd}$, which is found to stay largely at a constant $\sim 1.2$ in microgravity while an additional term inversely proportional to impact velocity arises in the presence of gravity. The constant term can be understood from momentum transfer along the penetration direction while the additional term suggests the influence of internal stress built-up due to gravity. Similar discrepancy is also found for the initial peak of the drag force. This analogy provides novel insights into the nature of granular drag in microgravity and sheds light on future space missions.

\end{abstract}

%\keywords{drag force, drag coefficient, granular materials, microgravity}%Use showkeys class option if keyword
                              %display desired
\maketitle

Walking on the beach, we experience the drag of sand under our feet, and very similarly, NASA’s Perseverance rover experience it while driving on Mars. Along with recent developments in space science and exploration,
~\cite{Wurm2021, Ballouz2021, Feng2022}, there is a growing interest in understanding the influence of gravity on granular drag. Two main motivations are the development of effectively operational space vehicles with reliable space aeronautics and the application of projectiles in asteroid impact mitigation studies~\cite{Ruiz-Suarez2013, Katsuragi2016, Wurm2021, Ballouz2021, Feng2022}. In this context, the most relevant physics question is to which extent the drag force laws known for simple liquids~\cite{Chhabra2006} can be connected to complex granular flows which are driven far from thermodynamic equilibrium~\cite{Jaeger1996}. For fluid drag, continuous investigations since the pioneer work of Poisson have led to practical ways to classify various flow regimes and to quantitatively estimate drag coefficient~\cite{Poisson1832, Michaelides2023}. For granular drag, specific tasks are to understand the dissipation mechanisms experienced by projectiles and rovers under microgravity conditions~\cite{Katsuragi2013, Clark2014, Kang2018,  Takehara2014, Katsuragi2018}, and to identify the most significant characteristic length and time scales of these processes~\cite{Altshuler2014, Sunday2022}.

In experiments, various non-invasive experimental techniques have been introduced to track projectiles~\cite{Huang2020, Kostler2021, Schroeter2022} on granular beds. Based on the measurements, empirical laws and theoretical models of granular drag have been proposed for inertial and quasi-static regimes~\cite{Katsuragi2013, Clark2014, Katsuragi2016, Meer2017, Kang2018}. Driven by the widespread applications such as pile driving and impact cratering~\cite{Ruiz-Suarez2013, Wurm2021}, general models on fluid-like drag forces have been introduced, evolving from Poncelet's ballistic analysis to unified drag force model that incorporate depth-dependent terms~\cite{Katsuragi2007, Katsuragi2016, Meer2017, Jing2022}. More elaborated approaches that include the influence of interstitial air, packing density, cohesion on granular drag, as well as conditions for infinite penetration of the projectile, reminiscent to terminal velocity in a fluid, have been explored~\cite{Ciamarra2004, Umbanhowar2010, Pacheco2011, Royer2011, Marston2012, Kondic2012}. Despite the progress, challenges persist in recent space missions~\cite{Ballouz2021, Spohn2022, Espinosa2023} due to the lack of complete understanding of the various contributions of granular drag and couplings in between, as well as the complex rheological behavior of granular media~\cite{Goldman2008, Roth2021, Jing2022}.

\begin{figure}[t]
\includegraphics[width=0.95\linewidth]{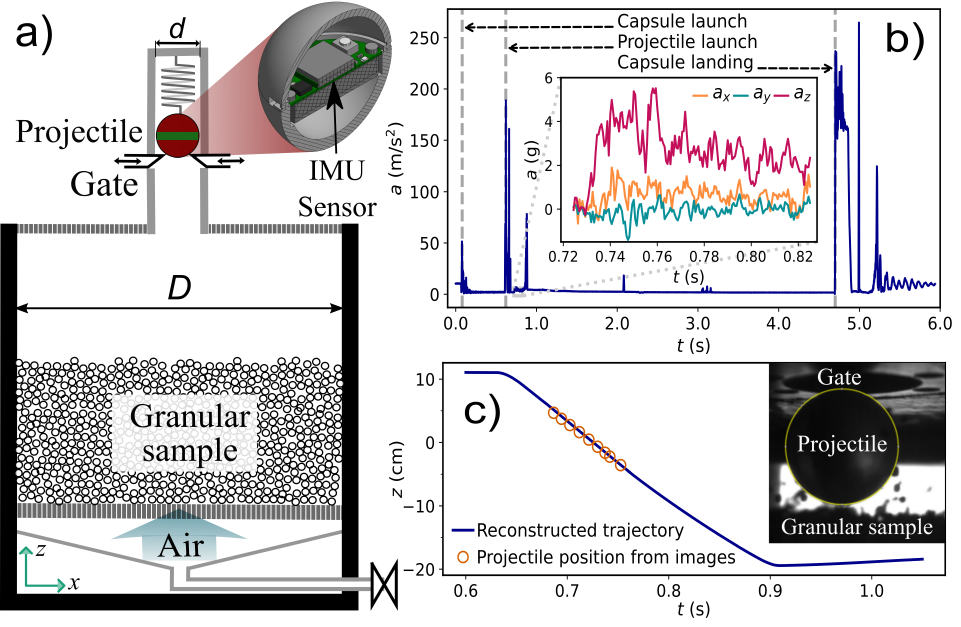}
\caption{(a) Droptower experimental Setup with the 3D printed spherical projectile with embedded IMU sensor highlighted. (b) Representative raw acceleration data collected from the IMU sensor during one capsule drop. Inset of (b) magnifies the onset of projectile impact on granular sample in microgravity. (c) A reconstructed trajectory of the projectile using the sensor data along with the projectile positions obtained from side view images, as illustrated in the inset.}
\label{fig:setup}
\end{figure}

In this Letter, we demonstrate that inertial term dominates granular drag in microgravity, leading to exponential decay of projectile velocity with penetration depth. The cavity generated behind the projectile exhibits a cone shape with a certain opening angle, in contrast to the convection induced cavity collapsing in earth gravity~\cite{Royer2005, Huang2020}. Consequently, the dimensionless granular drag coefficient defined similar to fluid drag is used to characterize granular drag as a function of gravity and impact velocity. The analysis of granular drag through the dimensionless drag coefficient helps us understand the nature of the drag force and the corresponding influence of gravity, thus pave the way for a first-principles understanding of granular drag.  

Figure~\ref{fig:setup} shows a sketch of the setup used in the experiment campaign at the Bremen Droptower (ZARM), which provides an excellent microgravity condition of $10^{-6}g_{\rm E}$ with $g_{\rm E}\approx \qty{9.81}{m/s^2}$ gravitational acceleration on earth~\cite{Kampen2010}. The setup is composed of a cylindrical polyvinyl chloride (PVC) container with inner diameter $D=31.0$\,cm and height $H=21.5$\,cm. Before each experiment, expanded polypropylene (EPP) particles of a certain mass are added to the container till filling height of $15$\,cm. More specifically, Neopolen P9230 or P9270~\cite{BASF2018} with bulk densities $\rho_{\rm b}\approx \qty{30}{kg/m^3}$ or $\qty{70}{kg/m^3}$ and diameter $d_{\rm p} \approx 3.4 \pm 0.4$\,mm or $2.7 \pm 0.3$\,mm are used in different experiments, respectively. Before each capsule drop, the granular sample is air fluidized to provide reproducible initial packing density $\phi\approx 0.60$, which is also used in simulations. Side view images are captured with a high speed camera (Phantom FASTCAM MC2). A loaded spring mounted on top of the container ejects the projectile into the granular medium at specific initial velocities $v_0$. A pre-calibrated IMU sensor board (Mbientlabs, Metawear C) is embedded in the 3D printed spherical projectile with a diameter of $d=32$\,mm and mass $m_{\rm i} = 15.72$\,g. Acceleration $\vec{a}$ and angular velocity $\vec{\omega}$ vectors of the sensor are recorded with a sampling rate of $1600$\,Hz and subsequently transmitted via Bluetooth to the on-board controller (Raspberry PI). Because $\vec{a}$ and $\vec{\omega}$ are measured in the sensor coordinate system, post-processing involving stepwise coordinate transformation and integrations are needed to reconstruct a three-dimensional trajectory of the projectile~\cite{Kostler2021}.

The representative raw accelerometer data shown in Fig.\,\ref{fig:setup}(b) illustrate three transition times (gray dashed lines): (i) Start of microgravity triggered by capsule release, measured by the transition from $a = g_{\rm E}$ to $0$ with $ a \equiv | \mathbf{a} |$. The granular sample expands slightly ($\sim 5\%$) at the beginning because of the elastic potential release. (ii) Launch of projectile by the release of the spring, which is set to be $\qty{0.5}{s}$ after capsule drop. Subsequently, $\vec{a}$ starts to fluctuate as the projectile moves. When the projectile touches the granular layer after the free-flying period [inset of Fig.~\ref{fig:setup}(c)], acceleration in the vertical direction $a_{\rm z}$ starts to increase, signifying the influence of granular drag. Fig.~\ref{fig:setup}(c) shows the reconstructed trajectory from the IMU sensor, along with the position data captured by analyzing the side-view images through Hough transformation~\cite{Kimme1975}. Subsequent peaks indicate abrupt events such as collisions with the container bottom. (iii) Landing of the capsule in the deceleration container, which is also filled with Styrofoam particles.

Control experiments are conducted in the lab with rescaled gravitational acceleration $\tilde{g} \equiv g / g_{\rm E}=1$ using the same experimental setup and protocol. Moreover, discrete element method (DEM) simulations with Hertz-Mindlin model~\cite{Huang2020} are employed to explore granular dynamics associated with the impact. In the simulations, initial packing of 410,240 spheres with random radii ranging from $\qtyrange{1.6}{1.8}{mm}$ are generated under $\tilde{g}=1$ and $0.1$. Subsequently, gravity is removed to allow sample relaxation similar to the experimental condition in microgravity. The numerical conditions, such as the coefficient of restitution $e$, Young's modulus of the particles, are set to match the experiments with more details described in a previous work~\cite{Huang2020}. To acquire the continuum fields from DEM simulations, coarse-graining methodology with a gaussian kernel size of $w \equiv \overline{r}/2 = \qty{0.85}{mm}$ is used~\cite{Goldhirsch2010,Huang2020}.

\begin{figure}[t] %{htbp}
    \centering
    \includegraphics[width=0.95\linewidth]{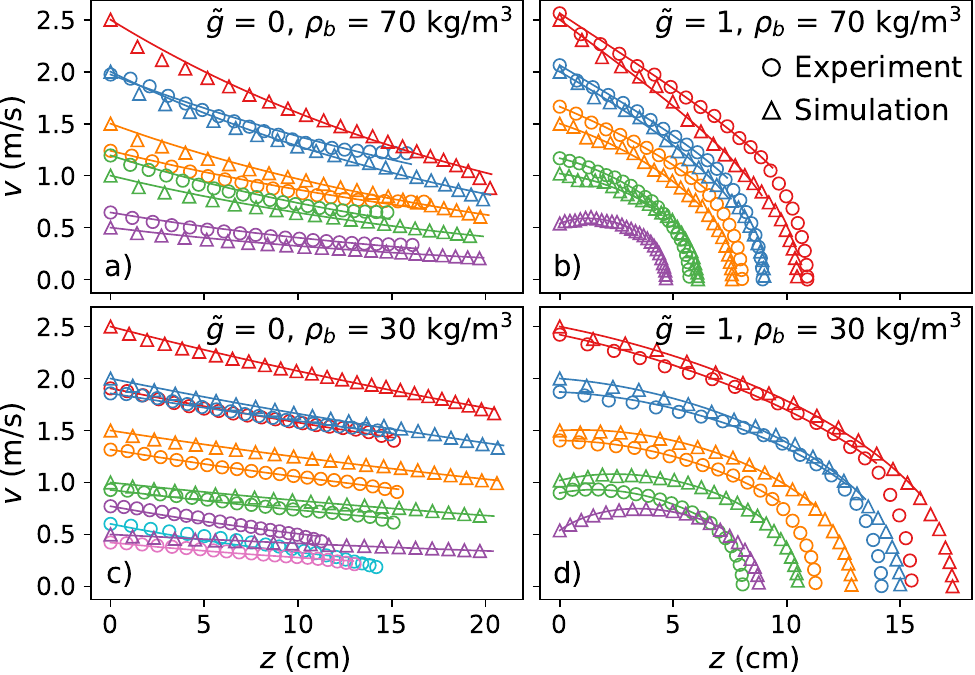}
   \caption{A comparison of vertical velocity $v$ vs penetration depth $z$ curves under micro- (left panels) and normal (right panels) gravity conditions. Spherical and triangle symbols correspond to experimental and numerical results. Solid curves denote fits to the data. See text for more details.}
    \label{fig:vz}
\end{figure}

Figure~\ref{fig:vz} compares projectile trajectories in micro- ($\tilde{g}=0$) and normal ($\tilde{g}=1$) gravity conditions for both samples. Qualitatively, the projectile always settles at a certain depth for $\tilde{g}=1$, but this is typically \emph{not} the case in microgravity shown in the left panels: The projectile penetrates through the granular sample and bounces with the container bottom in all cases except for the three experimental runs with lighter P9230 particles and lowest $v_0=\qty{0.42}{m/s}$, $\qty{0.60}{m/s}$, and $\qty{0.73}{m/s}$ (lower left panel). Moreover, $v$ always decays monotonically with $z$ and the decay curves are qualitatively different from $\tilde{g}=1$: For $\rho_{\rm b} = \qty{30}{kg/m^3}$, velocity decays with $z$ in the range from $z=0$, defined as the height at which $a_{\rm z}$ starts to deviate from 0 (i.e., granular drag starts to take effect), to the point where the intruder comes to rest or to the height $d/2$ above the container bottom to avoid boundary effect~\cite{Stone2004}. All data within the range are used for follow-up analysis of granular drag coefficient. Simulation data in microgravity extend to larger $z$ because of the larger bed height. Note that microgravity experiments with $v_0 = \qty{1.9}{m/s}$ and lighter particles are conducted twice and the results agree quantitatively well with each other, showing the reproducibility of the experiments. Due to the slight expansion of granular bed at capsule launch, the initial velocity $v_0$ in simulations cannot always match those in the experiments. Nevertheless, overall agreements between experimental and numerical results are found except for the three experimental runs mentioned above. The existence of a threshold velocity $v_{\rm c}$ between $\qty{0.73}{m/s}$ and $\qty{0.93}{m/s}$ suggests additional energy dissipation term, which might be attributed to enhanced internal frictional force that is independent on gravity or stronger influence of electrostatic forces, etc. 

%Note that $f = \kappa L \left(1-e^{-z/L}\right)$ with $L$ a characteristic length scale is also used in literature considering its saturation at larger $z$ following Jassen's law~\cite{Katsuragi2007, Pacheco2011, Katsuragi2016, Huang2020}.

%Note that considering the saturation of $f$ at larger $z$, following Janssen's law~\cite{Katsuragi2007, Pacheco2011, Katsuragi2016, Huang2020}, yields very comparable results. 

Quantitatively, we analyze the data with
$m_{\rm i} \dot{v} = m_{\rm i}g-\gamma v^2 - f$,
which includes the inertial term and a frictional `hydro-'static drag $f=\kappa z$ with parameters $\gamma$ and $\kappa$, respectively.  As shown in Fig.\,\ref{fig:vz}, the empirical law captures projectile trajectories well for both $\tilde{g}$ conditions.
Note that in the limiting case, $\tilde{g} = 0$,
the depth-dependent term can be safely ignored because the force chains are largely destroyed at the beginning of the capsule release. Thus, we have $m_{\rm i} \dot{v} =-\gamma v^2$, which explains the qualitative difference of $v$ vs.~$z$ curves with or without gravity. Consequently, the velocity of the projectile can be estimated analytically via $v=v_0 e^{-\frac{\gamma}{m_{\rm i}}z}$. It shows that the projectile impact will not stop moving until $z \to \infty$ for cohesionless particles in microgravity. In reality, it will collide with the container bottom as we observe in most experimental and all simulation runs. 

\begin{figure}[h]
    \centering
    \includegraphics[width=0.85\linewidth]{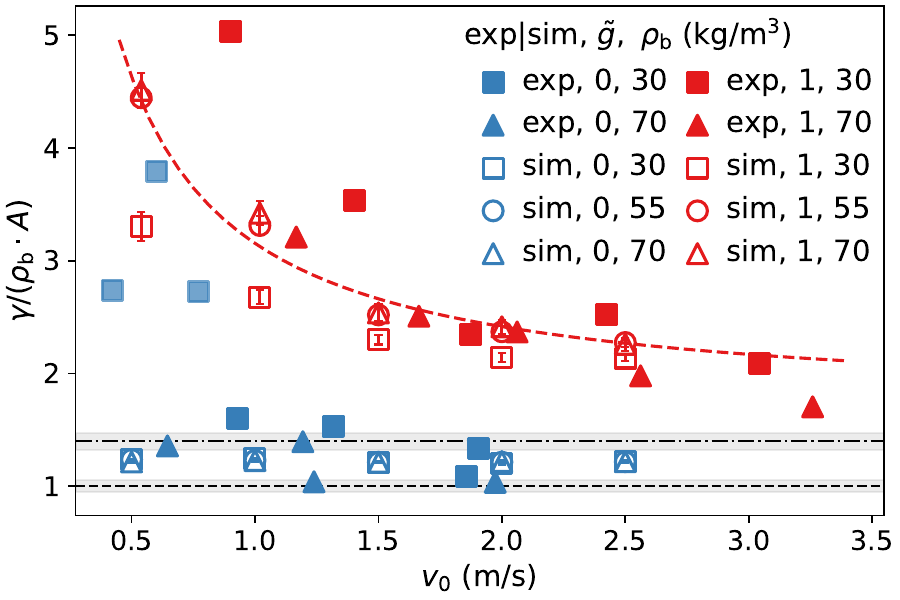}
    \caption{Granular drag coefficient $C_{\rm gd} \equiv \gamma/(\rho_{\rm b} A)$ as a function of initial impact velocity $v_0$ for different $\tilde{g}$ and $\rho_{\rm b}$.}
    \label{fig:drag}
\end{figure}

\begin{figure} %{htbp}
    \centering
    \includegraphics[width=0.95\linewidth]{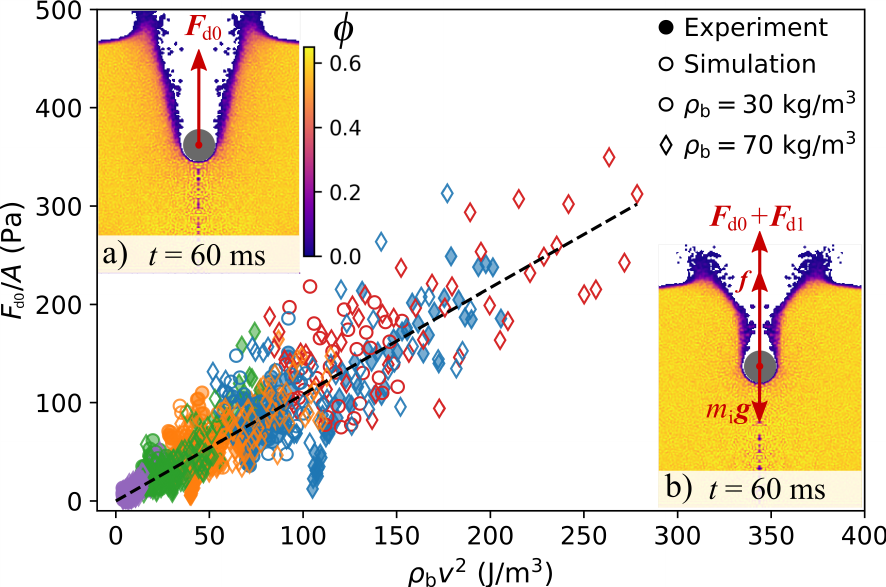}
    \caption{Effective stress $F_{\rm d0}/A$ as a function of $\rho_{\rm b}v^2$ for all $\tilde{g}=0$ data. The dashed line corresponds to a linear fit to all the data. The insets show the simulation results of the packing density field $\phi$ of a granular sample ($\rho_{\rm b}=70$ kg/m$^3$) responding to the impact of the projectile with $v_0=2.5$\,m/s in micro- (a) or normal gravity (b).}
    \label{fig:fv}
\end{figure}

Given the dominating role of inertial drag, granular drag coefficient is defined as $C_{\rm gd} \equiv \gamma / (\rho_{\rm b} A)$ with $A = \pi d^2/4$ cross section area of the projectile, reminiscent of fluid drag. Figure~\ref{fig:drag} shows the dependence of $C_{\rm gd}$ on $v_0$ for both gravity conditions. Interestingly, $C_{\rm gd}$ is found to stay close to a value slightly above unity in microgravity ($1.2\pm0.2$ for experiments and $1.22\pm0.01$ for simulations), except for the three experimental runs with $v_0 < v_{\rm c}$ and lighter P9230 particles (shaded in light blue). As the projectile moves through granular media, continuous momentum transfer from the projectile to the grains occurs with the momentum loss of the intruder and the gain of granular media being $\mathrm{d}\mathbf{p_{\rm i}} = m_{\rm i} \mathrm{d}\mathbf{v}$ and $\mathrm{d}\mathbf{p_{\rm gm}} = \rho_{\rm b} A \mathrm{d}l \cdot \mathbf{v}$ with the stepwise penetration depth $\mathrm{d}l \approx v \mathrm{d}t$, respectively. If we assume, quite ideally, that all momentum being transferred along the direction of $\mathbf{v}$ and the granular particles being pushed forward are confined in a cylinder of diameter $d$ and length $v \mathrm{d}t$, momentum conservation $\mathrm{d}\mathbf{p_{\rm i}}+ \mathrm{d}\mathbf{p_{\rm gm}}= 0$ yields $C_{\rm gd}=F_{\rm d0}/(\rho_{\rm b} A v^2) = 1$ with drag force under microgravity $F_{\rm d0}\equiv \mathrm{d}p_{\rm i}/\mathrm{d}t$ inertia dominated granular drag force. In reality, $C_{\rm gd}$ is slightly larger than one for two possible reasons: (i) Inelastic collisions of the projectile with interacting granular clusters of average size $N_{\rm c}$ and normal coefficient of restitution $e \approx 0.4$~\cite{Huang2020}, following Clark \textit{et al}.\,\cite{Clark2014}. Considering $N_{\rm c} \approx 1$ for cohesionless particles in microgravity due to the absence of gravity induced force chains, we have $C_{\rm gd} \approx 1+e$. As a guide to the eyes, we show two shaded region centered at the dash-dot $1+e$ and dotted $1$ horizontal lines in Fig.\,\ref{fig:drag}. The width of the shaded region corresponds to the uncertainty of $\rho_{\rm b} \equiv \rho_{\rm p} \phi$ with particle density $\rho_{\rm p}$ and fluctuating packing density $\phi$. (ii) As revealed by numerical simulation (see insets of Fig.\,\ref{fig:fv}), there exists momentum transfer along the radial direction, leading to a cone shaped cavity with a certain opening angle $\alpha\approx 21^\circ$, which grows slightly with $\rho_{\rm b}$ and $v_0$. Considering the additional factor $\tan{(\alpha/2)}$, we have $C_{\rm gd} \approx 1.2$. Note that the range of drag coefficient is in agreement with previous work~\cite{Katsuragi2007,Katsuragi2013,Sunday2022}. However, the above analysis indicates that gravity, as well as the frictional angle, does not play an essential role in $C_{\rm gd}$.  

In presence of gravity, re-organization of weight bearing force chain network along with collapsing of the cavity plays additional roles as the projectile penetrates granular media, leading to additional contributions to granular drag coefficient shown in Fig.\,\ref{fig:drag}.
$C'_{\rm gd}$, the granular drag coefficient with gravity, decays monotonically as $v_0$ grows. An average of the fits of all $\tilde{g}=1$ data sets with $C'_{\rm gd} = k v_0^{-1}+C_{\infty}$ yields $k = 2.1$ and $C_{\infty} = 1.3$, which is shown in Fig.\,\ref{fig:drag} as dashed curve. The connection between micro- and normal gravity conditions suggests that the inertial drag term is \emph{not} completely independent of gravity. $C_{\rm gd} \sim 1.2$ in microgravity sets the baseline for the inertial drag term $\gamma v^2$ due to momentum transfer, while the presence of gravity contributes to an additional drag term with coefficient $C'_{\rm gd} \propto 1/v_0$, which is reminiscent of the drag coefficient of fluids at low Reynolds number $\rm Re$. Although $\rm Re$ for granular media is not well-defined given the complex rheological behavior, velocity dependent drag coefficient was found in previous work~\cite{Bruyn2004,Huang2020,Jing2022}. In practice, the $1/v_0$ dependence suggests a way to estimate granular drag coefficient $C'_{\rm gd}$ with $k$ and $C_{\infty} \sim C_{\rm gd}$, which determines granular drag coefficient in microgravity.

The above comparison reveals the role of gravity on the $v_0$ dependent inertial drag, which can be further rationalized as follows. In addition to momentum transfer induced by the projectile penetrating through granular media, the shear force $F_{\rm d1}=\tau A$ with $\tau$ effective shear stress between the projectile and surrounding particles should also be considered. According to $\mu(I)$ rheology~\cite{MiDi2004}, $\tau$ can be predicted with $\mu(I)P$, where the inertial number $I \equiv \dot{\gamma} d/\sqrt{P/\rho_{\rm b}} $ with $\dot{\gamma}$ local shear rate and $P$ normal stress applied at the interface between the projectile and the surrounding granular media. With the presence of gravity, enhanced force chains between particles in contact results in higher $P$ and consequently higher shear force acting on the projectile. In addition, higher impact velocity leads to higher $I$, which lead to larger $\tau$. Therefore, we speculate that the gravity dependent inertial drag term arises from shear force induced by the friction between the projectile and the granular media during penetration, which is in addition to the $F_{\rm d0}$ term due to momentum transfer. Further investigations are needed to quantitatively characterize $C_{\rm gd}$ from first-principles.

\begin{figure}[t]
    \centering
\includegraphics[width=0.85\linewidth]{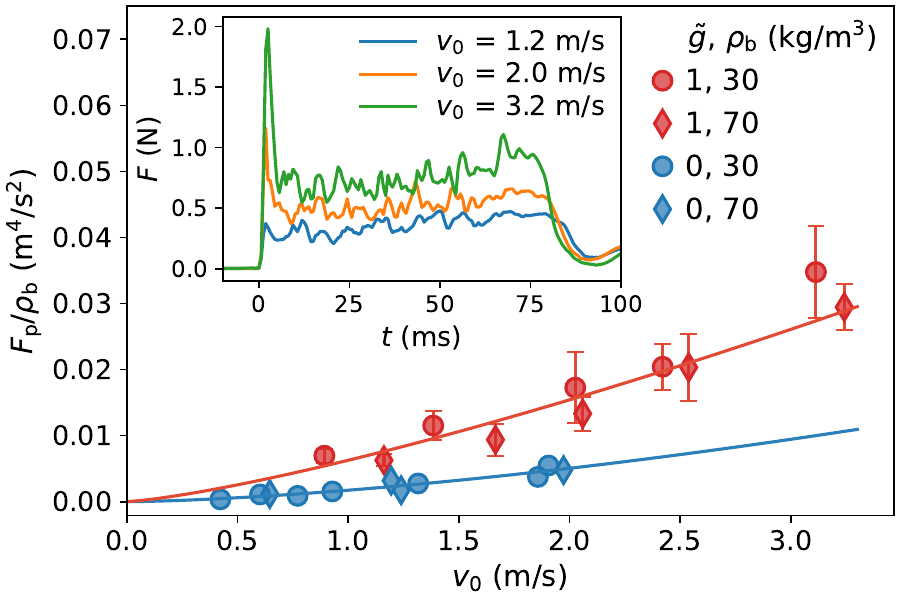}
    \caption{Influence of gravity on the rescaled initial peak of $F_{\rm p}$ as a function of $v_0$ for the experimental results. Inset shows a sample drag force measured for $\rho_{\rm b} = \qty{70}{kg/m^3}$ and $\tilde{g}=1$. The solid curves correspond to power law fits of the data sets with exponents $1.3 \pm 0.1$ and $1.6\pm0.4$ in normal and microgravity, respectively.}
    \label{fig:fp}
\end{figure}

Focusing on the microgravity condition, granular drag force can be represented in the effective stress $F_{\rm d0} / A$ versus energy density $\rho_{\rm b} v^2$ plane in Fig.\,\ref{fig:fv}. Despite starting from various $v_0$, all data fall onto a line with zero offset with a slope of $\approx 1.2 \pm 0.2$ and $\approx 1.12 \pm 0.02$ for experimental and simulation results, respectively. For relatively high $v_0$, there exists a force peak (see inset of Fig.\,\ref{fig:fp}) at the initial stage of the impact. To avoid the influence of the peaks, only data collected after delay time $20$\,ms are used in the fitting. 

The density field from numerical simulations (see the insets of Fig.\,\ref{fig:fv}) reveals another qualitative difference between microgravity and normal impacts: In contrast to the field at $\tilde{g}=1$\,\cite{Huang2020}, the cavity generated by the impact does not collapse as time evolves. Instead, it expands as a cone with an opening angle of $\alpha \sim 21$ degrees behind the projectile, similar to a Mach cone with a fixed ratio between front propagation speed $v_{\rm f}$ and traveling speed of the projectile $v_{\rm f}/v = \tan{(\alpha/2)} \approx 0.2$. Moreover, the density fields also show that only particles in a limited region are being disturbed as the projectile penetrates, supporting the above assumption of small $N_{\rm c}$.

From an application perspective, it is important to quantify the initial force peak $F_{\rm p}$ under microgravity conditions (see Fig.\,\ref{fig:fp}). In fact, $F_{\rm p}$ typically defines a threshold value below which penetration fails\,\cite{Zacny2008, Feng2022, Spohn2022}. As shown in the inset of Fig.\,\ref{fig:fp}, $F_{\rm p}$ arises shortly after the projectile touches the granular surface. 
The force then reaches a plateau with fluctuations arising from collisions with surrounding particles before gradually settling. In agreement with previous investigations, as $v_0$ increases, the force values increase systematically, suggesting the role of inertial drag. Fig.\,\ref{fig:fp} examines the dependence of $F_{\rm p}$ as a function of $v_0$. Remarkably, when rescaling the $F_{\rm p}$ experimental data by $\rho_{\rm b}$, the data points collapse for the same $\tilde{g}$, and are in good agreement with power law relation $F_{\rm p} = w (v_0/v_{\rm b})^{\beta}$ with $w$ and $\beta$ fit parameters. The presence of gravity leads to larger $F_{\rm p}$ and factor $w$. Qualitatively, the granular particles are not yet mobilized at the time of initial impact, thus the projectile interacts with a larger cluster of particles (i.e., larger $N_{\rm c}$ following previous discussion) and consequently leading to larger peak for $\tilde{g} = 1$. In microgravity, the transition $\tilde{g} \to 0$ effectively weakens $F_{\rm p}$ from the collisions. Quantitatively, the fits yield exponents $\beta=1.3 \pm 0.1$ and $\beta=1.6\pm0.4$ under normal gravity and microgravity, respectively. Although the range of exponent is in agreement with previous work\,\cite{Krizou2020, Mandal2024,Sunday2022,Roth2021}, a higher $\beta$ in microgravity is found than that in normal gravity. Further experiments with a wider range of $v_0$ are needed to understand the influence of gravity on the initial force peak.

To summarize, this work demonstrates that granular drag in microgravity condition is dominated by inertial drag $F_{\rm d0} = -\gamma v^2$, which leads to qualitatively different dynamics such as infinite penetration. Importantly, the dimensionless drag coefficient is found to be a constant of $1.2 \pm 0.1$, which can be understood with momentum transfer. This value sets the baseline for the inertial drag term, which includes an additional term $\propto 1/v_0$ in the presence of gravity. Moreover, the cavity generated by the projectile forms a cone-shaped cavity with a fixed opening angle. This investigation paves the way for quantitative predictions of granular drag coefficient towards zero gravity, and provides additional insights on the inertial drag term in the empirical drag force law. Further investigations on the initial force peak and the evolution of the cavity will shed light on a more comprehensive understanding of granular drag based on first principles, which can not only help with future space missions, but also serve as a bridge between Newtonian and non-Newtonian fluids on drag forces in general.

%\section{Acknowledgment}

This work is partly supported by the Deutsches Zentrum f\"ur Luft- und Raumfahrt e.V. (DLR) through granting the droptower experimental campaign at ZARM, and by the Startup Grant from Duke Kunshan University. The Collective Dynamics Lab at Duke Kunshan University is partly sponsored by a philanthropic gift. RC Hidalgo acknowledges financial support from MINCIN Spanish Government Grant PID2023-146422NB-I00 supported by MICIU/AEI/10.13039/501100011033. We thank Klaus \"Otter, Andreas Forstner, Thorben K\"onemann, Chen Lyu and Changwen Pan for technical support and calibrating the experimental system. Inspiring discussions with Ingo Rehberg and Matthias Schr\"oter are gratefully acknowledged. 

\bibliography{drag}

\end{document}